\documentclass[preprint,amscd,amsmath,amssymb,verbatim,showpacs]{revtex4}

\usepackage{graphicx}

\begin{document}

\title{Black Holes and the Strong CP Problem}

\author{John Swain}
\affiliation{Department of Physics, Northeastern University, Boston, MA 02115, USA}
\email{john.swain@cern.ch}
\date{revised version June 21, 2010}

%%%%%%%%%%%%%%%%%%%%%%%%%%%%%%%%%%%%%%%%%%%%%%%%%%%%%%%%%%%%%

\begin{abstract}
\section*{\bf Summary}
The strong CP problem is that $SU(3)$ gauge field instantons naturally
induce a CP violating term in the QCD Lagrangian which is constrained
by experiment to be very small for no obvious reason. We show that
this problem disappears if one assumes
the existence of at least one black hole somewhere in the universe. 
The argument is reminiscent of Dirac's argument for the quantization of charge,
in which the existence of one monople anywhere in the universe suffices to require the
quantization of electric charge everywhere.
\end{abstract}
%\maketitle

\maketitle

\clearpage

%%%%%%%%%%%%%%%%%%%%%%%%%%%%%%%%%%%%%%%%%%%%%%%%%%%%%%%%%%%%%
\section{Introduction}

In quantum chromodynamics (QCD)\cite{generalrefs} - the generally 
accepted theory of quarks and gluons -  there was a prediction that there
should be a light pseudoscalar particle associated with the conserved current
associated with global chiral rotations of the quarks. 
No such meson was observed, and this was called the
``U(1) problem''. It was realized that the quantum effects spoiled 
the conservation of the quark axial current, making its divergence
proportional to $TrF_{\mu\nu}F^{\ast\mu\nu}$
where $F_{\mu\nu}$ is the $SU(3)$ field strength, $F^{\ast\mu\nu}$ its Hodge dual,
and the trace is taken over $SU(3)$ indices. 

This divergence corresponds to a CP-violating Lagrangian density of the form

\begin{equation}
L_{CP-violating} = \frac{\theta g^2}{32\pi^2}TrF_{\mu\nu}F^{\ast\mu\nu}
\label{eqn:CP}
\end{equation}

\noindent where $g$ is the $SU(3)$ gauge coupling constant, and
$\theta$ is a free parameter. Overall, this expression is proportional to
the Pontryagin density, which on integration over spacetime
is an integer topological charge representing the number $n$ of times that $S^3$ (considered
as physical space ${\mathbb{R}}^3$ plus a point at infinity) nontrivially
``winds around'' $SU(3)$. The physical gauge-invariant vacuum is 
constructed as a superposition of states of winding
number $n$, each weighted by $e^{in\theta}$ with the sum running from
$n=-\infty$ to $n=\infty$ in order
to preserve gauge invariance under the ``large'' gauge transformations which
are not continuously connected to the identity. $\theta$ is not determined by
the theory, and can, in principle, take any value between $0$ and $2\pi$.

When the weak interactions and quark masses are included,
$\theta$ is shifted by an amount
$arg(det(m))$ where $m$ is the quark mass matrix, but the basic form
of the expression remains the same and unless the shifted $\theta$ is zero (or $\pi$, but this
subtlety will not concern us here) 
this term leads to a (CP-violating) electric dipole moment $|d_n|$ for the neutron.
The present upper bound $|d_n|<2.9\times 10^{-26}e\cdot$cm, where
$e$ is magnitude of the electron charge\cite{neutron-edm} , 
implies $\theta <10^{-9}$. The puzzle of why $\theta$ is so small is 
the ``strong CP problem''.

A wide variety of solutions have been
proposed, generally involving new physics. 
Many postulate particles
called axions\cite{generalrefs} associated with an additional $U(1)$ symmetry
which can be used to rotate $\theta$ to zero. These have not been observed and
are in general quite constrained by astrophysical considerations.
Other ideas include adding dimensions to the usual
3+1 that we know \cite{stdiml}, or making them
some fractional value a little less than four \cite{stdimf}. 
Two-dimensional fundamental
objects (2-branes) \cite{branes} have been considered, as have
microscopic wormholes \cite{worm},  hypothetical
new interactions \cite{newint}, new (non-axion) particles \cite{newpart},
supersymmetry \cite{SUSY}, and magnetic monopoles\cite{monopCP}.
It has also been claimed that certain choices of regularization techniques
could solve the problem\cite{reg}. It has even been suggested that
a staggering $10^{32}$ standard model copies could do the job
\cite{gf}. It has also been argued\cite{Fort} that the strong CP problem
might naturally not appear at all if one simply reformulated QCD in terms of
holonomies  (gauge invariant traces of Wilson loops).
This list of ideas and references is not meant
to be complete, but rather to show that the problem has
driven theorists to a wide range of quite exotic scenarios
in the search for an explanation. 
Despite all this creativity, the strong
CP problem is still generally considered unsolved.

The point of this paper is that the problem could be resolved without unobserved
exotica, and without spoiling
the solution of the $U(1)$ problem, if the spacetime integral of the Pontryagin density were
somehow to be zero -- something I now argue will happen if one allows for the existence
of even one black hole.

In elementary particle physics one usually ignores gravity, and works with
quantum field theory in flat and topologically trivial spacetime. 
While quantum field theory in a general curved spacetime\cite{qftc} is
highly nontrivial, the question asked here only requires a little topology. 

First let us recall where the instantons come from that lead to the strong CP
problem\cite{Jackiw-and-Rebbi,CDG}. We look for $SU(3)$ gauge
field configurations $A_\mu$ which  go to the identity (up to a gauge transformation)
at spatial infinity, with all the directions at infinity identified. These turn out to
fall into topologically distinct classes labelled by elements of $\pi_3(SU(3))$.

For completeness, and to make clear the origin of
the $\pi_3(SU(3))$, let us repeat the argument in more detail.
Pure gauge configurations are of the form
$A_\mu=iU^\dag(x)\partial_\mu U(x)$ where $U(x)$ takes its values in
$SU(3)$ and $x=(t,\vec{x})$. Using the gauge freedom to set $A_0=0$, which essentially means we
consider time-independent fields, only partially fixes the gauge. If we require
$U(\vec{x})=1$ as the spatial $\vec{x}$ goes to infinity in all directions
this is the same as looking for maps from spatial ${\mathbb{R}}^3$ compactified
at infinity (that is, $S^3$) into $SU(3)$.
Instantons and
correspond to homotopy classes $[S^3,SU(3)]$ of these maps.
By definition, $[S^3,SU(3)]=\pi_3(SU(3))$ and since $\pi_3(SU(3))=Z$ we have
homotopically distinct maps labelled by the integers, which turn out to be
the very topological charges that come from the integration of the Pontryagin
density in equation \ref{eqn:CP}.

The key observation of this paper is that if we have black holes present and
repeat the argument, we
should replace $[S^3,SU(3)]$ by $[M,SU(3)]$ where $M$ is a manifold (now
with boundary) created from the $S^3$ described above with a 3-ball bounded
by a 2-sphere excised for each black hole present -- effectively we are
removing a set of distinct points (and balls around them) from space.

Physically, we require that the gauge fields go to the identity
(up to gauge invariance)
on the surfaces of black holes (as well as at infinity), in 
a similar spirit to reference \cite{CraneSmolin} in which this condition is invoked to argue 
for spacetime foam as a universal regulator.
Note that we don't need to assume spacetime
foam or wormholes (as have been used to argue for solutions to the strong CP
problem before) and the black holes in question
need not be virtual or microscopic -- any astrophysical black holes 
(or, indeed just one) would do. In particular, $\pi_1(M)$ is assumed to be
trivial, as is usual in considerations of the strong CP problem (and for
which there is no experimental evidence to the contrary) . In many ways,
this is meant to be a very conservative solution to the strong CP problem
invoking essentially no new physics beyond what is generally known.

Now let us consider the homotopy classes of maps $\left[M,SU(3)\right]$ 
from $M$ to $SU(3)$. $M$ is clearly simply connected ($\pi_1(M)=0$), as
every closed loop can be continuously shrunk to a point. If we consider
possibly topologically nontrivial maps from $M$ to $SU(3)$ then the
usual Postnikov construction \cite{topology} tells one that one has to 
consider $\pi_2(SU(3))$, but this is zero, and one is left with nothing
to worry about except $\left[M,K(\pi_1(SU(3)),1)\right]$
with $K(\pi_1(SU(3)),1)$ being the relevant Eilenberg-MacLane space.

By definition that means that $\left[M,K(\pi_1(SU(3)),1)\right]=H^1(M,\pi_1(SU(3)))$.
Since $M$ is simply connected, one immediately sees that this is zero, and
thus all maps from $M$ to $SU(3)$ are homotopically trivial (continuously
deformable to the identity). We could argue directly
that it is also zero due to the
fact that $SU(3)$ is simply connected and $\pi_1(SU(3))=0$.

If one wants to argue that the
true gauge group should be $SU(3)$ with its $Z_3$ center divided
out\cite{Oraif}, making the first
homotopy group nonzero, then the first argument given 
in the above paragraph still makes the case. 

The integral in equation \ref{eqn:CP}
now vanishes as the corresponding topological charge is zero, and
the strong CP problem would seem to be solved. Clearly, analogous
arguments hold for any finite-dimensional Lie group $G$ in place of
$SU(3)$ since $\pi_2(G)$ is always zero in this case\cite{topology} and the same
reasoning applies. Some care is needed if multiply connected $M$ is
considered since one does not want to induce a $\theta$-like term
for the $U(1)$ of electromagnetism. Such an electromagnetic $\theta$ term
is absent in standard analyses
since $\pi_3(U(1))=0$ and thus there are no $U(1)$ instantons
to worry about.

In the case of topologically more complicated spacetimes additional
gravitationally-induced CP violating effects may be present\cite{Deser-Duff-Isham}.
In particular, terms proportional to $f_{\mu\nu}f^{\ast\mu\nu}$ where $f$ is the
electromagnetic field strength tensor and $R^{ab}_{\mu\nu} R^{\ast\mu\nu}_{ab}$
where $R$ is the spacetime curvature tensor can be present. These contributions
are not usually considered part of the ``strong CP problem'', although it is 
very interesting that these terms are not obviously suppressed by powers of
the Planck scale. In the case of the $R^{ab}_{\mu\nu} R^{\ast\mu\nu}_{ab}$
the spacetimes involved clearly are not of the form considered here since
corresponding instantons do not refer to topologically nontrivial gauge
fields {\em over} spacetime but rather topologically nontrivial spacetimes.
Whether the arguments made here can be extended to this case is not
obvious, but I hope to be able to return to this interersting question in future work.
Of related interest is also \cite{Seiberg} in which the suggestion is made that
the usual instanton sums may need to be modified in some theories.

As this paper was being completed, I became aware
of a related paper\cite{Etesi} by Etesi. This author considers
both black and ``white'' holes (which it is not clear exist), finding results
for $\left[M,SU(3)\right]$ which agree with those here. The claim
in that paper 
however is not that the Pontryagin term integrates to zero, but rather
that one should consider a sort of  ``effective homotopy'' which takes
into account the causal structure of the relevant spacetime and for which
the corresponding homotopy classes are not trivial and the strong
CP problem remains. The idea is that one should only consider
homotopies whose initial and final stages can be compared by an
observer in finite time. This leads to a re-appearance
of the $\theta$ vacuum structure which we just got rid of, and the
solution of the strong CP problem is
based an additional assumption which is certainly not required
in the usual formulation of the strong CP problem.
In fact $\theta$ arises in a quantum mechanical superposition of
states of all instanton numbers making even the meaning of a
suitable observer unclear at best. Indeed the term ``instanton'' refers to the fact
that one considers field configurations which can be thought of as 
at least approximately localized in time. This leads to that paper missing the key point
I make here which is
that even a single black hole (no ``white holes'' needed) suffices to
make all the $SU(3)$ field configurations topologically trivial. In this
way $TrF_{\mu\nu}F^{\ast\mu\nu}$ can still be nonzero locally to
solve the $U(1)$ problem, while globally all the corresponding gauge
field configurations are topologically trivial.

In contrast to essentially all other attempts to solve the strong CP
problem, the approach presented here requires no modification
of the standard treatment of the problem other than to include the
presence of normal (indeed classical) black holes as part of the
structure of spacetime. No undiscovered exotica need be invoked.

It may seem surprising that the existence of even one singular object - in this
case a black hole - could have implications for elementary particle physics,
but there is actually a rather old analog. Long ago in 1948,
Dirac had used topological arguments
to show\cite{monopoles} that the presence of just {\em one} magnetic monopole
would require electric charge everywhere to be quantized. Here we see that,
similarly, the presence of just one black hole can resolve the strong CP problem.

\section{Acknowledgements}

I would like to thank Ka\'{c}a Bradonji\'{c} and Tom
Paul for reading early
drafts of this paper. I would also like to thank Egil Lillestol, Danielle
Metral, Nick Ellis, and the hospitality
of the CERN Latin American School of High Energy Physics
2009 in Colombia where this work was started, Luis Alvarez-Gaum\'{e}
since it was during his lecture to the students that I started to think
about this problem again, and of course all the students who made
the school the great success that it was! I would also like to thank G\'{a}bor 
Etesi for email correspondence on the first draft of this paper, as well
as for pointing out that while his argument in reference \cite{Etesi} allowed for
white holes it does not require them, and for emphasizing the possible
differences in considering Euclidean and Minkowsi spacetimes in 
calcuations. Thanks are also due to Michael Duff and Stanley Deser for
email correspondence and 
bringing references \cite{Deser-Duff-Isham} and \cite{Seiberg} to my attention.
This work
was supported in part by the US National Science Foundation under grant
NSF0855388.

\end{document}